\documentclass[letterpaper]{article}
\usepackage{aaai}
\usepackage{times}
\usepackage{helvet}
\usepackage{courier}
\usepackage{graphicx}

\usepackage{xcolor}

\begin{document}
%
\title{Dynamic topic modeling of the COVID-19 Twitter narrative among U.S. governors and cabinet executives}
\author{Hao Sha \and Mohammad Al Hasan \and George Mohler\\
Indiana University - Purdue University\\ Indianapolis\\
\And P. Jeffrey Brantingham\\
University of California \\Los Angeles}

\maketitle
\begin{abstract}
\begin{quote}
A combination of federal and state-level decision making has shaped the response to COVID-19 in the United States. In this paper we analyze the Twitter narratives around this decision making by applying a dynamic topic model to COVID-19 related tweets by U.S. Governors and Presidential cabinet members. We use a network Hawkes binomial topic model to track evolving sub-topics around risk, testing and treatment.  We also construct influence networks amongst government officials using Granger causality inferred from the network Hawkes process.
\end{quote}
\end{abstract}

\section{Introduction} 

By mid-April 2020, the number of active COVID-19 cases has reached over 2 million and the 
number of deaths is over 140,000 world-wide. The United States has the largest share of confirmed cases (over 670,000) and confirmed deaths (over 27,000).  Without a vaccine 
yet available, states throughout the U.S. are attempting to control transmission and reduce strain on the healthcare system through school and business closings, along with shelter-in-place orders.  Careful planning and coordination is needed both to minimize risk from
the disease, and to minimize the long-term economic impact.  

In the U.S., a combination of federal and state-level decision making has shaped the country's response to COVID-19.
The response is quickly evolving, making it difficult to understand how decision
makers have influenced each other, and whom among the decision makers have emerged as leaders on different topics. To overcome this difficulty, we analyze the Twitter narrative of various decision makers through dynamic topic modeling.  Specifically, we analyze a dataset of all COVID-19 related tweets by U.S. Governors, the President, and his cabinet members
between January 1st 2020 and April 7th 2020. We use a Hawkes binomial topic model (HBTM) \cite{hbtm} to track evolving sub-topics around risk, testing and vaccination/treatment.  The model also allows for estimation of Granger causality \cite{xu2016learning} that we use to construct influence networks amongst government officials.

Our work contributes to the growing body of literature on social media analytics and COVID-19. A summary of the most related work is as follows. In \cite{cinelli2020covid}, general COVID-19 related topic diffusion across different social media platforms is analyzed.  In \cite{yin2020covid}, the authors study COVID-19 discussions on Chinese microblogs.  Gender differences in COVID-19 related tweeting is investigated in \cite{thelwall2020covid} and in \cite{thelwall2020retweeting} the authors analyze consensus and dissent in attitudes towards COVID-19. Geolocated tweets are used to estimate mobility indices for tracking social distancing in \cite{xu2020twitter}.

\section{Hawkes Binomial Topic Model}\label{sec_HBTM}
We analyze COVID-19 related tweets by U.S. governors and cabinet members using a network Hawkes binomial topic model\footnote{Code and data available at: https://github.com/gomohler/hbtm} (HBTM) \cite{hbtm} with intensity $\lambda_s(t,\vec{m})$ at node $s$ in the network determined by,
\begin{eqnarray}
&\lambda_s(t,\vec{m})=\mu_s(t) J_0(\vec{m}|p^s_{0})+&\label{G-HBTM}\\
&\sum_{t>t_i}\theta_{s s_i} \omega_{s s_i} e^{-\omega_{s s_i}(t-t_i)} J_1(\vec{m},\vec{m}_i |p^{s s_i}_{off},p^{s s_i}_{on}).\nonumber
\end{eqnarray}
A Hawkes process is a model for contagion in social media where the occurrence of a post increases the likelihood of more posts in the near future.  In the HBTM, tweets are represented as bags of words following a Binomial distribution.  When viewed as a branching process, the daughter event bag of words is generated by randomly turning on/off parent words through independent Bernoulli random variables.  
\begin{figure}[hbt]
\centering
\hspace{-.3cm}\includegraphics[scale=.4]{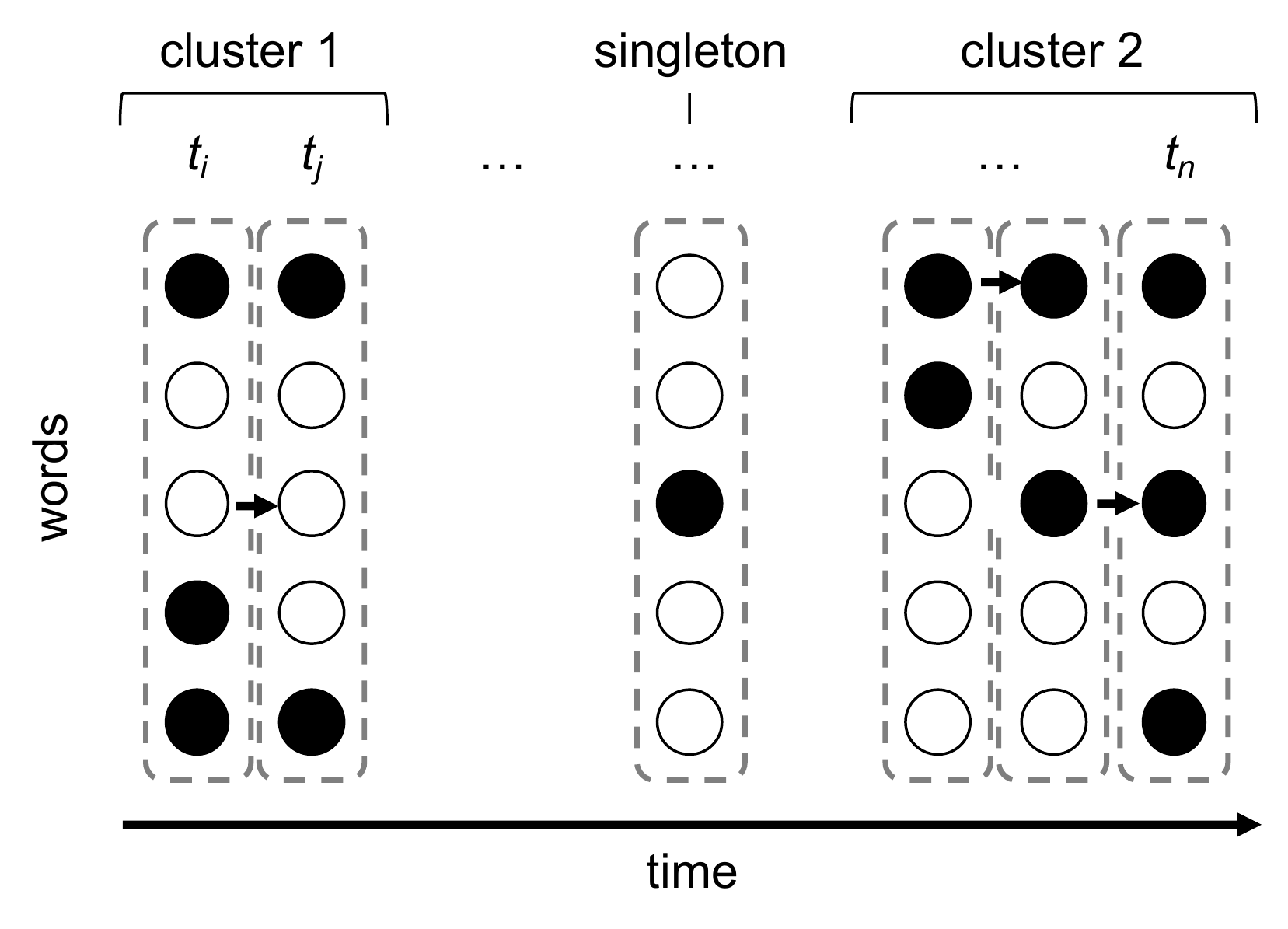}
  	\caption{\small In the HTBM, spontaneous events occur with marks generated by a binomial random variable over the dictionary of keywords contained in the data set.  Events then trigger offspring events whose marks are generated by switching parent event words off (white circle) with probability $p_{off}$ and on (black circle) with probability $p_{on}$. Unique events are delineated with dashed lines. Clusters are groups of parent daughter events connected by triggering.
}
  	\label{time_series}
\end{figure}

In Equation \ref{G-HBTM} events at time $t_i$ are associated with a mark $\vec{m}_i$, a vector of size $W$, the number of words in the overall dictionary across events.  The binary variables indicate whether each word is present or absent in the event at time $t_i$.  Spontaneous events occur according to a Poisson process with rate $\mu_s(t)$ at node $s$ in the network (here a node is either a governor or cabinet member).  Unlike in \cite{hbtm}, we let the spontaneous rate vary in time to reflect the exponential increase in overall COVID-19 related Twitter activity (for estimation we use a non-parametric histogram).  The mark vector of spontaneous events is determined by, 
\begin{equation}
J_0(\vec{m}|p^s_0)={p^s_0}^{\sum_{j=1}^W m_j}(1-p^s_0)^{W-\sum_{j=1}^W m_j},
\end{equation} 
which is the product of W independent Bernoulli random variables with parameters $p^s_0$

The parameter $\theta_{ss'}$ determines the expected number of tweets by individual $s$ triggered by a tweet by individual $s'$ and can be viewed as a measure of influence.  The expected waiting time between a parent-daughter event pair is given by $\omega_{ss'}^{-1}$.  The mark of a daughter event is determined by two independent Bernoulli processes.  Each word absent, or ``turned off," in the parent bag of words is added to the bag of words of the child event with probability $p^{ss'}_{on}$.  Each word present in the parent bag of words is deleted with probability $p^{ss'}_{off}$.  Thus $J_1$ is given by,
\begin{eqnarray}
&J_1(\vec{m},\vec{m}_i |p^{ss'}_{off},p^{ss'}_{on})=\\
&(p^{ss'}_{on})^{W^{\vec{m},\vec{m}_i}_1}(1-p^{ss'}_{on})^{W^{\vec{m},\vec{m}_i}_2}(p^{ss'})_{off}^{W^{\vec{m},\vec{m}_i}_3}(1-p^{ss'}_{off})^{W^{\vec{m},\vec{m}_i}_4},\nonumber
\end{eqnarray} 
where $W^{\vec{m},\vec{m}_i}_1$ is the number of words present in the child vector and absent in the parent vector, $W^{\vec{m},\vec{m}_i}_2$ is the number of words absent in both vectors, $W^{\vec{m},\vec{m}_i}_3$ is the number of words in the parent vector absent in the child vector, and $W^{\vec{m},\vec{m}_i}_4$ is the number of words present in both vectors.

After removing stop words we restrict the dictionary to the $W$ most frequent words, on the order of several hundred most frequent words across tweets. 
The Model given by Eq. \ref{G-HBTM} can be viewed as a branching process and is estimated using Expectation-Maximization (EM) \cite{hbtm}.  Using the EM algorithm for estimation has the added benefit that branching probabilities, estimates of the likelihood that tweet $i$ was triggered by tweet $j$, are jointly estimated with the model:
\begin{equation}
q_{ij}=\frac{\theta_{s_i s_j} \omega_{s_i s_j} e^{-\omega_{s_i s_j}(t_i-t_j)} J_1(\vec{m}_i,\vec{m}_j |p^{s_i s_j}_{off},p^{s_i s_j}_{on})}{\lambda(t_i,\vec{m}_i)}.\label{em1}
\end{equation}
These branching probabilities can then be clustered to generate families of dynamic topics over time \cite{hbtm}.

\subsection{Related work}
We note that Hawkes branching point processes in general are a popular model for mimicking viral processes on social media.  Previous studies have utilized temporal point processes to model Twitter  \cite{zhao2015seismic,simma2012modeling}, Dirichlet Hawkes processes \cite{du2015dirichlet,xu2017dirichlet,lai2014topic}, joint models of information diffusion and evolving networks \cite{farajtabar2017coevolve}, Hawkes topic modeling for detecting fake retweeters \cite{dutta2020hawkeseye}, and Latent influencers are modeled in \cite{tan2018indian} using an Indian buffet Hawkes process.  For a review of point process modeling of social media data see \cite{kim2020real}. 

\begin{figure}[h]
\centering
\includegraphics[width=0.45\textwidth]{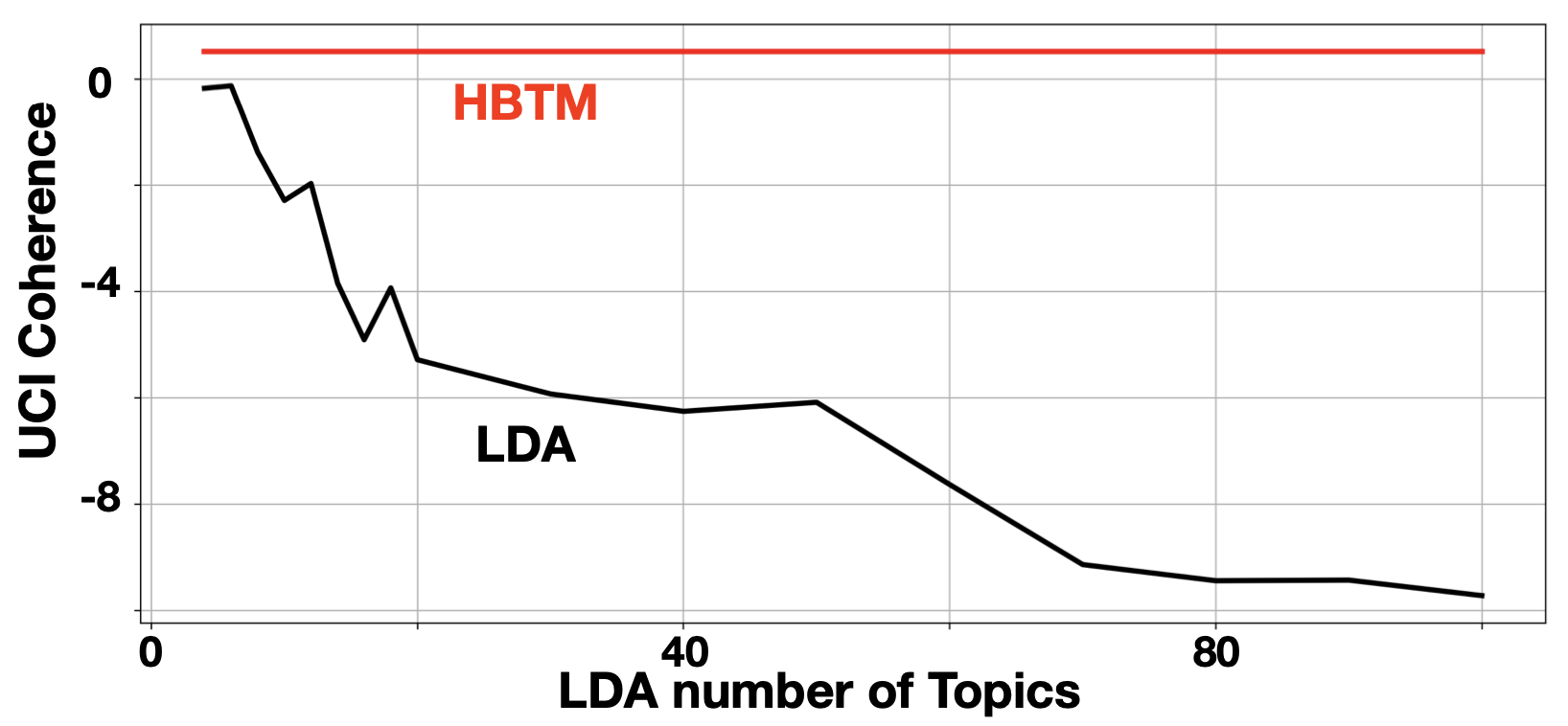}
\caption{UCI coherence of HBTM vs. LDA when applied to COVID-19 related tweets by governors and cabinet members.}\label{fig_coherence}
\vspace{-0.2in}
\end{figure}

Compared to standard LDA-type Hawkes processes, the HBTM has the advantage that it jointly estimates a network that can be used to measure influence; additionally, HBTM automatically detects the number of clusters.  The temporal aspect of HBTM-like dynamic topic models tend to improve topic coherence in relation to LDA (see Figure \ref{fig_coherence}).

\section{Data}
\begin{figure*}
\centering
\includegraphics[width=0.9\textwidth]{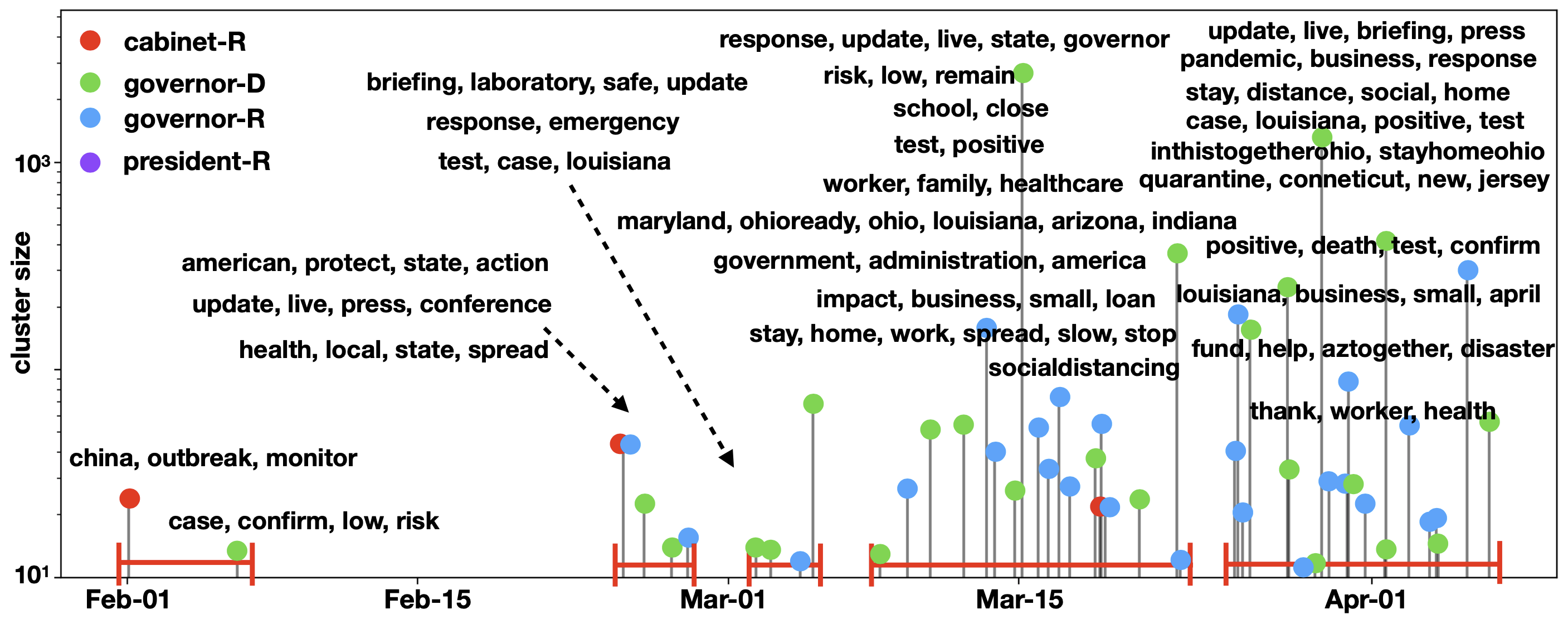}
\caption{Topic timeline. Clusters with size greater than $10$ are pinned. Keywords indicate the topic of the clusters. The marker color indicates the dominant component of the cluster.}\label{fig_topic_timeline}
\vspace{-0.2in}
\end{figure*}

We first collected the verified Twitter handles of all U.S. state governors, presidential cabinet members, and the president (a total of $73$ politicians, see Fig. \ref{fig_user_influcence} for their handles). Next, we used the Twitter API to query all tweets by these users during the period of January 1, 2020 to April 7, 2020. We then performed a keyword expansion \cite{buntain2018sampling,hbtm} to extract a list of keywords related to COVID-19. This method iteratively adds keywords to a query list whose frequencies in the set of matching tweets are significantly higher than in the general sample.  We then scanned the corpus with the expanded keyword list, obtaining a set of $7881$ COVID-19 related tweets by these politicians.  These tweets were further sorted in time-ascending order and converted to a bag-of-word representation.  The vocabulary was then restricted to the top 425 words according to frequency.


\section{Results}

 We cluster the data into space-time topics by sampling the branching probabilities $q_{ij}$ in Equation \ref{em1}.  In particular, we assign tweets to the same group when a link between tweet $i$ and tweet $j$ is sampled.  In Fig. \ref{fig_topic_timeline}, we show topic clusters over time consisting of more than $10$ tweets. Each marker height represents the size of the cluster and the most frequent keywords per marker indicate the topics of the clusters. 

The clusters show roughly four phases in time, with a significant gap between the first phase and the rest. In the first phase (early February), the federal government (most frequent handle @SecAzar, Alex Azar, Sec. of Health) informed the public of the \textbf{outbreak} in \textbf{China} and claimed to closely \textbf{monitor} the situation. Also in this phase several state governors (most frequent handle @NYGovCuomo, Andrew Cuomo, Gov. of New York) started reporting \textbf{confirmed} \textbf{cases}, but stated that the \textbf{risk} was \textbf{low}, as the number of cases was limited. 

The second and the third phases (early March) appeared almost a month later. From the keywords in these two phases, we can see that the government started to take \textbf{action} to \textbf{protect} the \textbf{American} citizens (possibly overseas in the regions of the outbreak). We can also see that \textbf{live updates} and \textbf{press conferences} were given to \textbf{brief} the public.  Keywords like \textbf{spread} and \textbf{emergency} indicate that the outbreak was getting worse in the U.S.  Meanwhile, the keyword \textbf{test} was mentioned frequently alongside \textbf{laboratory}, as limitations in U.S. testing was driving some of the narrative. 
\begin{figure}
\centering
\includegraphics[width=0.45\textwidth]{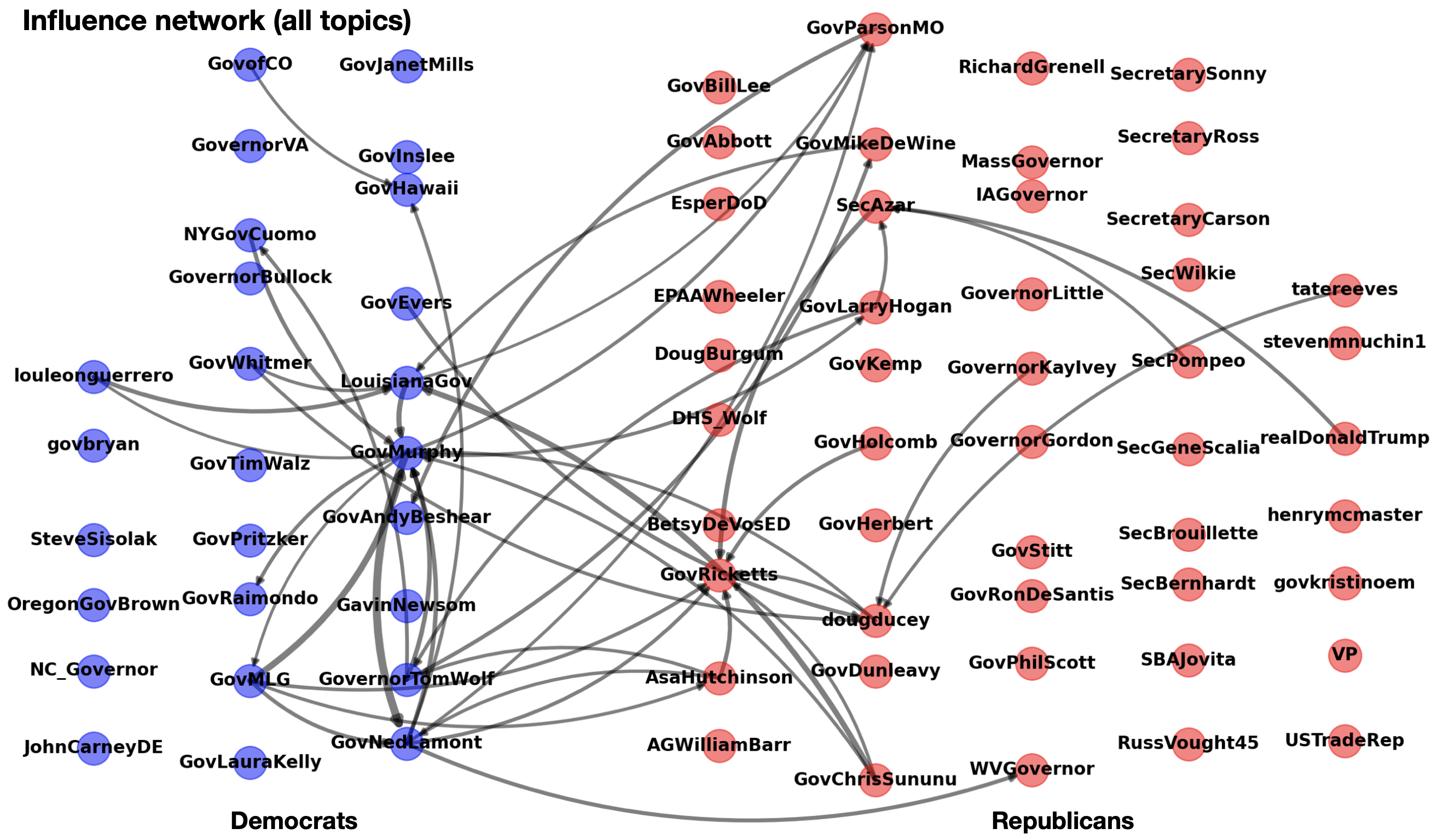}
\caption{Granger causality \cite{xu2016learning} influence network. Democrats (blue), Republicans (red). Weights of the edges of the directed graph correspond to the fraction of events estimated to be triggered across the edge. Edges with weights less than 10 are removed.}\label{fig_influence_network}
\end{figure}

\begin{figure*}[t]
\centering
\includegraphics[width=1.0\textwidth]{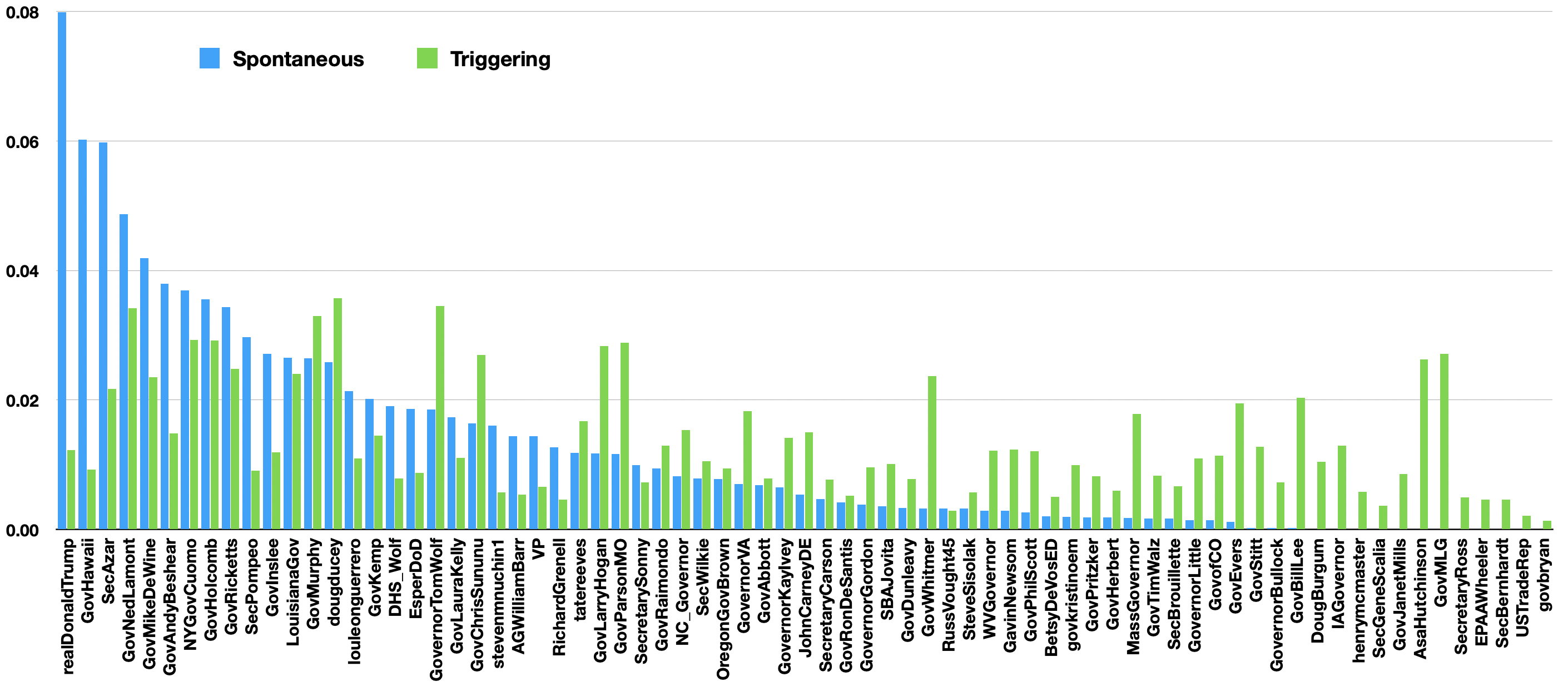}
\caption{Spontaneous vs. triggering effects of politicians on Twitter. Vertical axis: base intensities (spontaneous) and effective influences (triggering) are normalized over politicians; horizontal axis: Twitter handles of politicians. To save space, vertical axis is truncated at $0.08$, rendering President Trump's spontaneous rate off the chart ($\sim 0.16$).}\label{fig_user_influcence}
\end{figure*}

The fourth phase starts around mid-March, when clusters became larger and denser.  In this phase, \textbf{live updates} were held by many governors on a regular basis (the highest peak in Fig. \ref{fig_topic_timeline}). We also see the separation between the federal and state governments, as the clusters divided into \textbf{government, administration, america} and the various states (\textbf{maryland, ohio, louisiana, arizona, indiana}).  The Louisiana governor John Bel Edwards (@LouisianaGov) and the Ohio governor Mike DeWine (@GovMikeDeWine) were among the most active on Twitter sending information to the people in their respective states. 

The topic of \textbf{risk} appears in this phase, and the message is that risk \textbf{remains low}.  New topics also emerged on social distancing policies such as {\bf school close}, {\bf stay home},  and {\bf work (from) home}.  During the third phase the government began addressing problems like \textbf{healthcare} for \textbf{workers} and \textbf{families}, and \textbf{loan}(s) for \textbf{small businesses} due to the \textbf{impact} of the pandemic.  The slogan \textbf{socialdistancing} was widely adopted in this phase. 

In the most recent phase, a cluster with frequent words {\bf live update}, {\bf press conference}, and {\bf briefing} is the largest, alongside a narrative around the number of \textbf{tested}, \textbf{confirmed positive} and \textbf{death cases} in different states.  The Louisiana and Ohio governors continued to be the most active.  Also {\bf small businesses} remained a concern during this phase and the keyword {\bf disaster} indicates the negative impact of COVID-19. Meanwhile, \textbf{quarantine} and \textbf{stay home} were encouraged and reiterated on Twitter.  The sacrifices of \textbf{health workers} were acknowledged (\textbf{thank}).

In Figure \ref{fig_influence_network}, we show inferred influence among governors and cabinet members by plotting a network where each edge weight from $i\rightarrow j$ is determined by the total estimated number of tweets triggered at node $j$ by tweets from node $i$.  The network shows influence across party lines, with Democrat governors {\bf GovNedLamont}, {\bf GovernorTomWolf}, {\bf GovMurphy} and {\bf LouisianaGov} highly connected with Republican governors {\bf GovRicketts}, {\bf GovLarryHogan} and {\bf GovParsonMO}.  We caution that this network captures Granger causality \cite{xu2016learning}, and does not control for confounding effects.  In Figure \ref{fig_user_influcence}, we plot the estimated baseline rate of spontaneous tweets per governor and cabinet member, along with each individuals estimated influence (average number of subsequent tweets in the network directly triggered by a Tweet).  Here we observe that President Trump has the highest rate of spontaneous tweets, followed by the Governor of Hawaii and Secretary Azar.  Governors Ducey, Wolf and Lamont are the largest estimated influencers.


\subsection{Risk, treatment and testing sub-topics}
\begin{figure*}[t]
\centering
\includegraphics[width=0.9\textwidth]{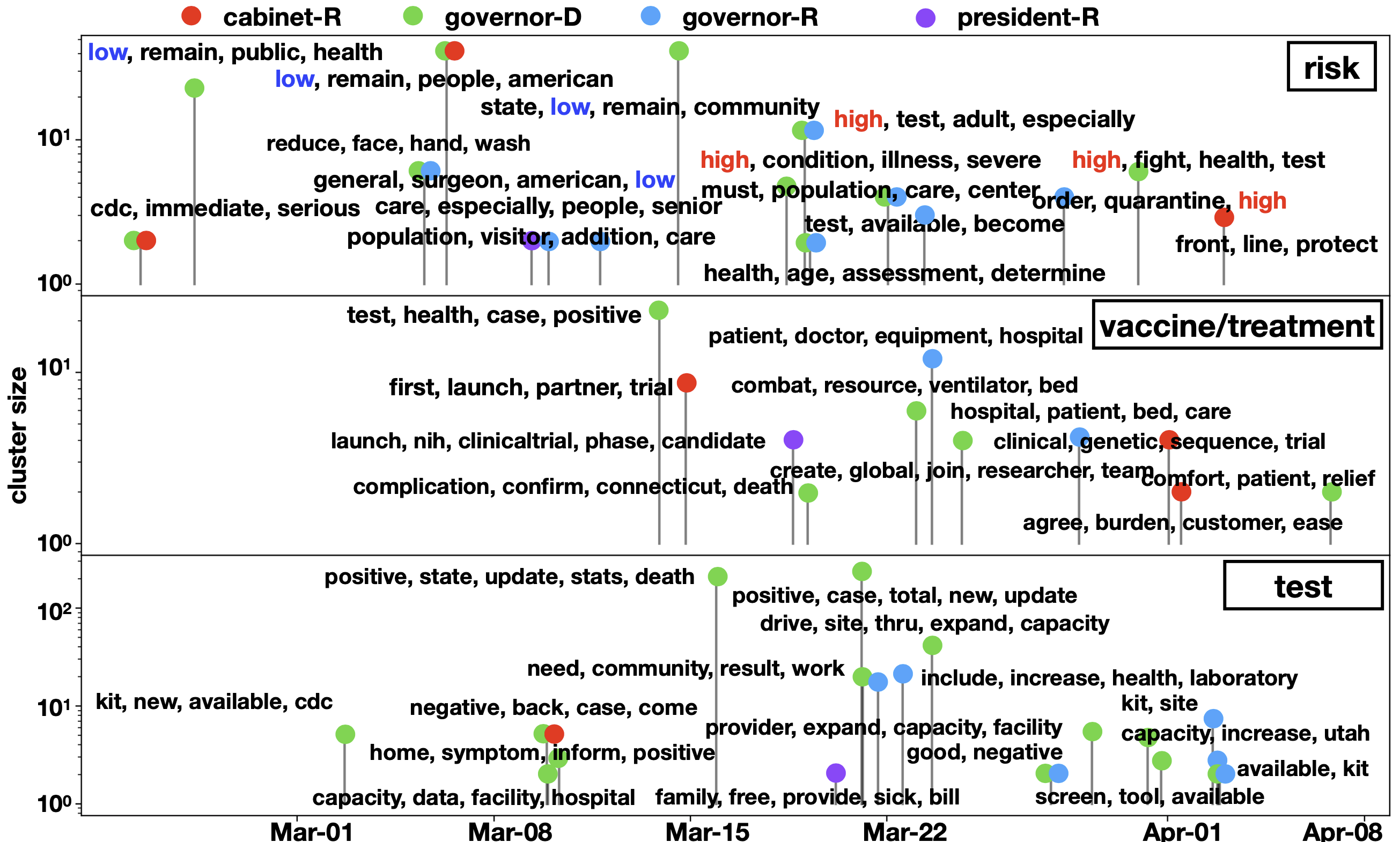}
\caption{Timeline of sub-topics on risk, treatment and testing. Clusters with size at least $2$ are pinned. Keywords indicate the topic of the clusters. The marker color indicates the dominant component of the cluster. }\label{fig_topic_timeline_three}
\end{figure*}

In addition to applying the HBTM to all COVID-19 related tweets, we also apply the model separately to three sub-categories. We first apply HBTM to tweets containing the word "risk". A sequence of clusters are illustrated in the top row of Fig. \ref{fig_topic_timeline_three}. The emergence of this sub-category coincides with the start of the second phase of the general timeline, and it appears that the \textbf{CDC} was among the first to mention how \textbf{serious} the risk was and asked for \textbf{immediate} actions. However, the subsequent clusters in early March indicate that both state and federal governments (Republicans and Democrats) were telling the public that the risk \textbf{remains} \textbf{low}. Also in this period, we observe calls for \textbf{washing hands} to \textbf{reduce} risk, and that \textbf{seniors} were identified to be the most vulnerable.  After March 15, the narrative changes and the \textbf{high} risk to the general \textbf{population} is acknowledged.  Keywords like {\bf age} and {\bf adult} indicate the high risk across \textbf{age} groups, even for young \textbf{adults}.  The word {\bf high} frequently co-occurs with {\bf test} and {\bf quarantine}; due to the \textbf{high} risk of transmission, state governments increased \textbf{testing} and enforced \textbf{quarantine}(s). Overall, from left to right, the sequence of clusters show a clear trend in the narrative from low risk in late February to high risk in April.

Next, we apply HBTM to tweets containing the words ``vaccine" and ``treatment". The resulting clusters are illustrated in the middle row of Fig. \ref{fig_topic_timeline_three}.  In mid-March, keywords {\bf launch}, \textbf{trial}, \textbf{clinicaltrial}, {\bf phase}, and {\bf candidate} indicate that vaccine \textbf{candidates} were identified and entered the \textbf{clinical trial phase}. We can also see the National Institute of Health (\textbf{NIH}) \textbf{partner} with the pharmaceutical industry in developing the vaccine.  Later in March, we start to see clusters where state governors (mainly Democrats) commented on the lack of \textbf{resources}, \textbf{equipment}, \textbf{ventilators}, and \textbf{hospital beds}. We also see cabinet members (specifically Sec. of Health @SecAzar) giving updates about vaccine development (\textbf{genetic sequence} and \textbf{clinical trial}).  Another narrative is around an agreement (\textbf{agree}) with insurance companies to \textbf{ease} the \textbf{burden} of the pandemic on their \textbf{customers}. Additionally, we see the request to \textbf{create global researcher team} in developing a vaccine. In general, the clusters here suggest that the search for a vaccine has been a collective effort that crosses political parties and national boundaries. 

\begin{table*}[h]
\caption{Officials ranked by in-degree (most influenced) and out-degree (most influential) in influence networks. }\label{table_ in_out_degree}
\centering
\begin{tabular}{lcc}
\hline
Topic & In-degree & Out-degree\\
\hline
all & \textcolor{blue}{GovMurphy}, \textcolor{red}{GovRicketts}, \textcolor{blue}{LouisianaGov} & \textcolor{blue}{GovNedLamont}, \textcolor{blue}{GovMurphy}, \textcolor{blue}{GovMLG}\\
risk & \textcolor{red}{GovMikeDeWine}, \textcolor{blue}{NYGovCuomo}, \textcolor{blue}{GovMLG} & \textcolor{red}{GovMikeDeWine}, \textcolor{blue}{GovPritzker}, \textcolor{red}{SecAzar}\\
treatment & \textcolor{red}{SecAzar}, \textcolor{blue}{GovNedLamont}, \textcolor{blue}{GovofCO} & \textcolor{blue}{GovofCO}, \textcolor{red}{GovChrisSununu}, \textcolor{blue}{GovNedLamont}\\
test & \textcolor{blue}{GovNedLamont}, \textcolor{red}{GovMikeDeWine}, \textcolor{blue}{LouisianaGov} & \textcolor{blue}{NYGovCuomo}, \textcolor{red}{GovHerbert}, \textcolor{red}{GovKemp}\\
\hline
\end{tabular}%
\end{table*}

In the bottom row of Fig. \ref{fig_topic_timeline_three}, we show clusters found by applying HBTM after filtering the dataset on the keyword ``test". In early March, we see that \textbf{new} test \textbf{kits} were \textbf{available}.  Tweets mention (\textbf{negative}) test results of some individuals by the Democrat governors and cabinet members.  Concern about the \textbf{capacity} of testing \textbf{facilities} and \textbf{hospitals} is also discussed in early March.  In mid-March, testing is expanded to the \textbf{community}, followed by requests for \textbf{expanding} \textbf{facility capacity} and \textbf{increasing laboratories}. During this period, state governors (especially Democrats, the two highest green markers in Fig. \ref{fig_topic_timeline_three}) start updating test results (in particular number of \textbf{positive cases}) and providing \textbf{stats} in their press conferences. The HBTM model identifies a cluster in which \textbf{drive thru site} is suggested as a way to \textbf{expand} testing \textbf{capacity}. In early April, we observe that the narrative has shifted away from a lack of testing resources; keywords indicate that \textbf{screen tools}, test \textbf{kits}, and test \textbf{sites} are available, and the testing \textbf{capacity} has \textbf{increased}.

\begin{figure}[t]
\centering
\includegraphics[width=0.45\textwidth]{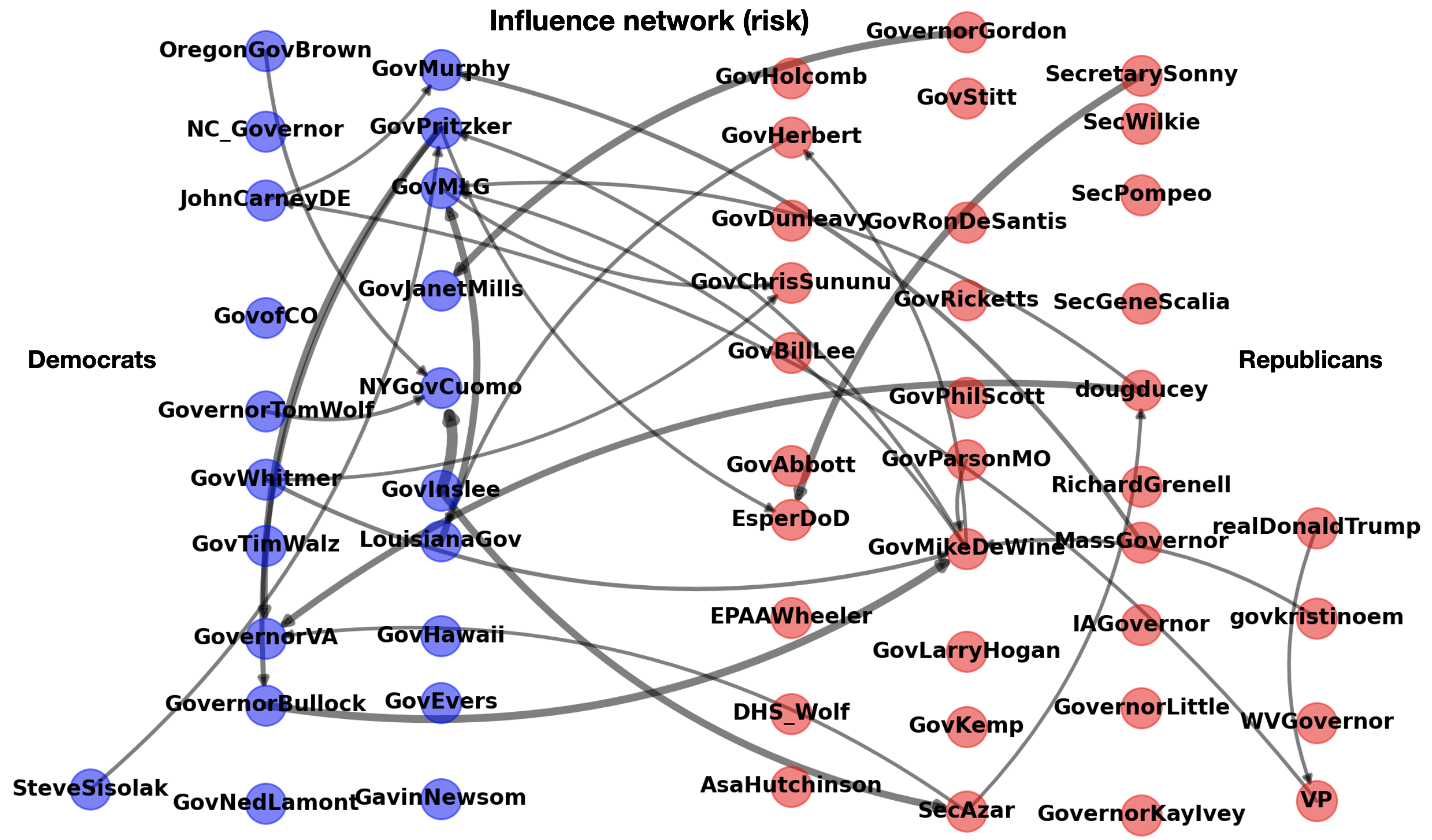}
\hspace{.5cm}
\includegraphics[width=0.45\textwidth]{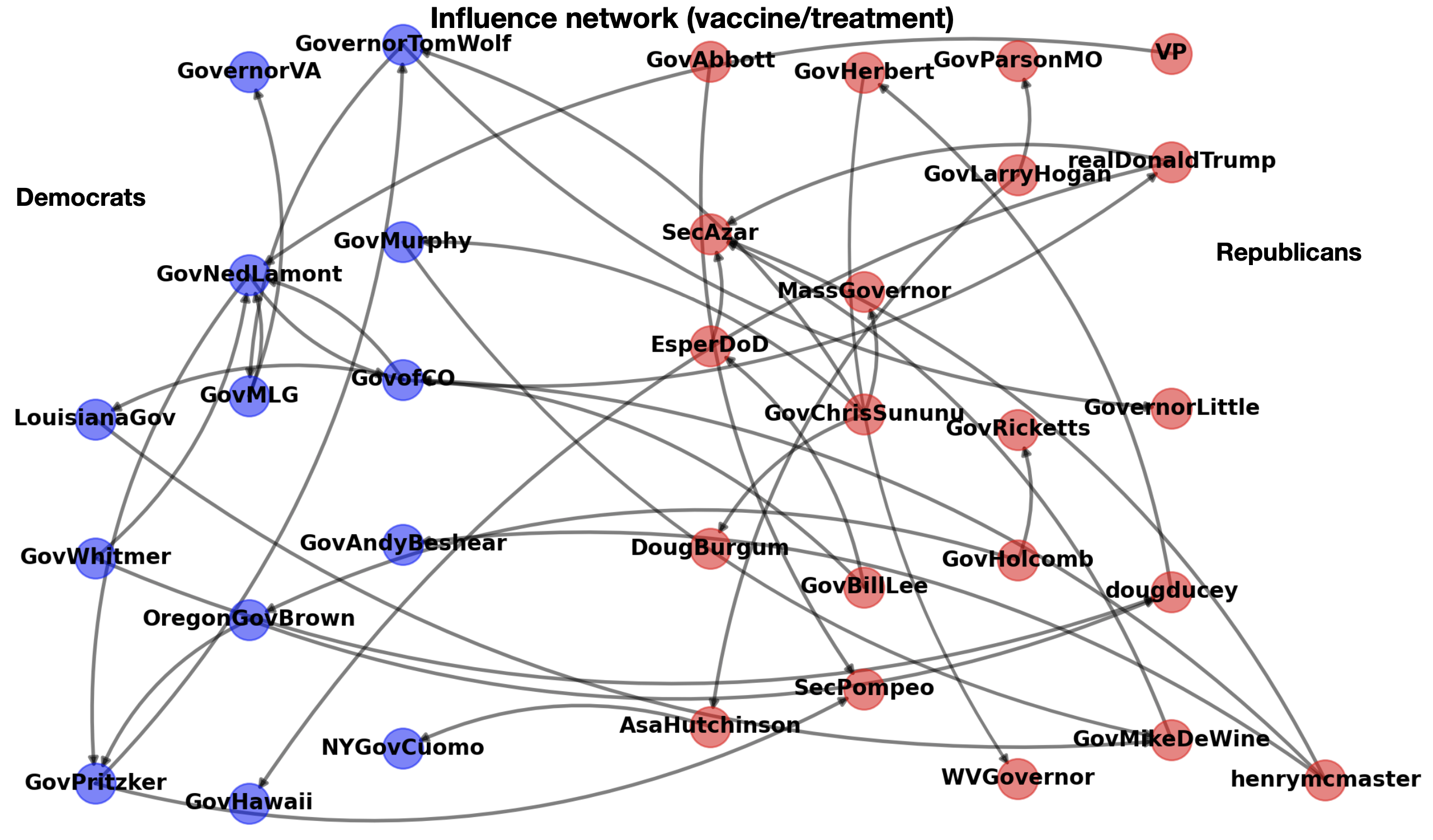}
\hspace{.5cm}
\includegraphics[width=0.45\textwidth]{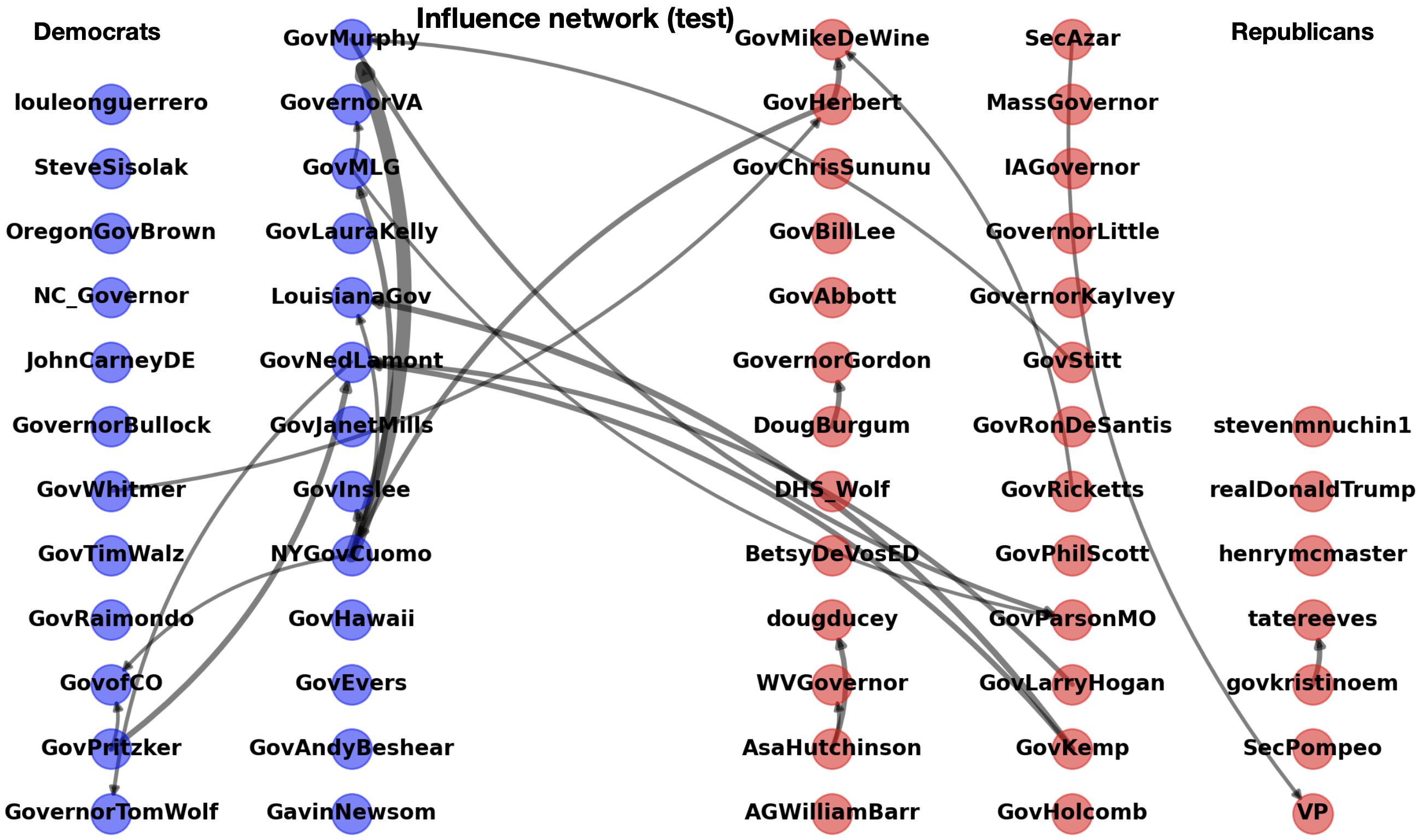}
\caption{Granger causality influence network for ``risk" (top), ``treatment" (middle) and ``test" (bottom) sub-topics.\label{fig_influence_network_sub_topics}}
\end{figure}

 In Figure \ref{fig_influence_network_sub_topics}, we plot Granger causality influence networks for the risk, treatment and testing sub-topics.  Again we see connections crossing party lines.  In the case of testing, the network is characterized by a dense set of connections between a select set of governors.  The risk and treatment networks are characterized by more active nodes with fewer connections.  In Table \ref{table_ in_out_degree} we also list the most influential officials by sub-topic along with those officials most influenced.



\

\section{Conclusion}
We analyzed the COVID-19 Twitter narrative among U.S. governors and presidential cabinet members using a Hawkes binomial topic model.  We observed several narratives between January 1st and early April 2020, including a shift in the assessment of risk from low to high, discussion of a lack of testing resources which later subsided, and sub-topics around the impact of COVID-19 on businesses, efforts to create treatments and a vaccine, and calls for social distancing and staying at home.  We also constructed influence networks amongst government officials using Granger causality inferred from the network Hawkes process.   President Trump stands out for spontaneity, yet appears to have little influence with respect to network cross-excitation.  Polarization is not obvious in the Granger influence networks; we observe a high level of cross party event triggering and influence seems more geographically clustered and related to state size.

We see several potential directions for future work.  Here we limited the analysis to only COVID-19 related tweets among U.S. government officials.  The HBTM can be used to explore the COVID-19 narrative among the general population and may highlight issues around trust in institutions, adherence to social distancing, and economic impacts.  Furthermore, analyzing non-COVID related tweets by government officials prior to the pandemic and constructing an evolving influence network may provide insights into how bi-partisan cooperation changes during national emergencies.


\bibliographystyle{aaai}
\bibliography{biblio,biblio_hbtm}

\end{document}